\documentstyle[multicol,osa,epsf]{revtex}
\begin{document}
\title{Dispersion-managed soliton in optical fibers with zero average dispersion}

\author{P. M. Lushnikov$^{1,2}$}
\address{
$^1$ Theoretical Division, Los Alamos National Laboratory, MS-B284, Los Alamos, New
Mexico, 87545
\\
$^2$ Landau Institute for Theoretical Physics, Kosygin St. 2, Moscow, 117334, Russia
   }

\maketitle

\begin{abstract}
The dispersion-managed (DM) optical system
with step-wise periodical variation of dispersion
is studied in the framework of
path-averaged Gabitov-Turitsyn equation. The soliton solution  is obtained by iterating the
path-averaged equation. The dependence of soliton parameters on dispersion
map strength is investigated together with the oscillating tails of
soliton.
\end{abstract}

~~~~~~~ {\it OCIS codes:}  060.2330, 060.5530, 060.4370, 190.5530, 260.2030.
\\

\begin{multicols}{2}

Dispersion-managed (DM) system \cite{kogelnik1}, which is the system with periodical variation of dispersion along optical fiber,
is one of the key point of current development of ultrafast
high-bit-rate optical communication lines \cite{nakazawa1,smithknox1,gabtur1,kumar1,mamyshev1,mollenauer1}. The main factor limiting the bit-rate
capacity is pulse-broadening due to the chromatic dispersion of the optical fiber which can be
overcome by periodical changing of sigh of fiber's dispersion to create very low (or even zero)
path-average dispersion. Lossless propagation of optical pulse in DM fiber is described by nonlinear
Schr\"odinger equation (NLS) with periodically varying dispersion $d(z)$:

\begin{equation}
i u_z +   d(z) u_{tt} +  |u|^{2} u =0,  \label{nls1}
\end{equation}
where $u$ is the envelope of optical pulse, $z$ is the propagation distance and all quantities are
made dimensionless. Consider two-step periodic dispersion map:
$d(z)=\langle d\rangle+\tilde d(z)$, where $\tilde d(z)=d_1$ for $0<z+n L<L/2$
and $\tilde d(z)=-d_1$ for $L/2<z+n L<L$, $\langle d \rangle$ is the path-averaged dispersion, $d_1$ is the
amplitude of dispersion variation, $L$ is a dispersion period and $n$ is an arbitrary integer number. Eq.
$(\ref{nls1})$  also describes pulse propagation in fiber with losses compensated by periodically
placed amplifiers if the distance between amplifiers is much less than $L$.

Provided a characteristic nonlinear length $Z_{nl}$ of the pulse is large:  $Z_{nl}\gg L,$ where $Z_{nl}=1/|p|^2$ and $p$ is a typical pulse amplitude,
the eq. $(\ref{nls1})$ can be reduced to path-averaged Gabitov-Turitsyn \cite{gabtur1} model:
\begin{eqnarray} \label{psiint1}
i \hat \psi_z(\omega) -\omega^2 \langle d \rangle \hat \psi + \frac{1}{(2\pi)^2}\int \frac{\sin{s\omega_1
\omega_2}}{s\omega_1 \omega_2 } \hat\psi(\omega_1+\omega)\times \nonumber \\ \hat\psi(\omega_2+\omega)
\hat\psi^*(\omega_1+\omega_2+\omega) d\omega_1 d\omega_2=0,
\end{eqnarray}
where $s=d_1 L/2$ is a dispersion map strength, $\hat\psi \equiv \hat u e^{i\omega^2\int^z_{L/4} \tilde d(z')dz'}$ and
$\hat\psi(\omega)=\int^\infty_{-\infty}\psi(t)e^{\imath \omega t}dt$ is Fourier component of $\psi$.
 Gabitov-Turitsyn model is based on the assumption of slow variation of $\hat\psi$ as a
function of $z$ on scales of the order of DM period $L$. This model is good supported by numerical
simulations \cite{turitsyn1,ablowitz1}. Returning to t-space one can get \cite{ablowitz1}:
\begin{eqnarray} \label{psiint1b}
i \psi_{ z} + \langle d \rangle \psi_{tt}- \frac{1}{2\pi s}\int Ci(\frac{t_1 t_2}{s})
\psi(t_1+t)\psi(t_2+t)\times \nonumber
\\ \psi^*(t_1+t_2+t) dt_1 dt_2=0,
\end{eqnarray}
where $Ci(x)=\int^x_\infty\frac{\cos{x}}{x}dx$ (note difference in
definition of $Ci(x)$ in comparison with \cite{ablowitz1}). It was found numerically \cite{smithknox1} that
Gaussian ansatz
\begin{equation} \label{psi0}
A_{Gauss}=p \exp{\big ( -\frac{\beta}{2}t^2\big )},
\end{equation}
where $p, \beta$ are real constants, is a rather  good approximation for the soliton solution
$u=A(t) e^{\imath\lambda z}$ ($A$ is real) of eq. $(\ref{nls1})$ for small and moderate values of
$t$ at space points $z=L/4+n L.$ Respectively $(\ref{psi0})$ is also a good approximation for the soliton solution
$\psi=A(t) e^{\imath\lambda z}$ of  $(\ref{psiint1b}).$
On soliton solution eq. $(\ref{psiint1b})$ takes the form:
\begin{eqnarray} \label{psiint1c}
-\lambda A + \langle d \rangle A_{tt}= \frac{1}{2\pi s}\int Ci\big(\frac{t_1 t_2}{s}\big ) A(t_1+t)\times   \nonumber
\\ A(t_2+t)A(t_1+t_2+t) dt_1 dt_2,
\end{eqnarray}

Based on this observation the soliton solution of $(\ref{nls1})$ was approximated in \cite{turmez1,kaup1} by Gaussian
exponent multiplied by a sum consisting of a finite number of Hermite polynomials. Because infinite set of
these polynomials is complete  in the space of square-integrable function,  $\psi$ can be expanded over this set.
But for large enough $t$ such approximation can not be effective because of highly oscillating tails \cite{ablowitz1,turmez1} of $A$ for
$t \to \infty$ which requires to take into account very large number terms of expansion to get
reasonable approximation. The oscillating tails of $A$ are of big importance because they are
responsible for interaction of a sequence of solitons launched in optical fiber which limits bit-rate
capacity \cite{mamyshev1}.

Here the convergent set of approximate solutions of $(\ref{psiint1c})$ is found by iterating that eq. for
$\langle d \rangle=0.$ Zero iteration $A^{(0)}$ is given by $(\ref{psi0})$.
 $-\lambda A^{(n)}$ is obtained by substitution of $A^{(n-1)}$ into right hand side of $(\ref{psiint1c})$ for $n=1,2, \ldots$
The oscillating
tails appear at first iteration already. Simultaneously one can get the dependence of $\beta$ in eq.
$(\ref{psi0})$ on dispersion map strength $s$.

Rewriting the kernel of
$(\ref{psiint1})$ via parametric integral:
\begin{equation} \label{paramc1}
\frac{\sin{s \omega_1\omega_2}}{s \omega_1\omega_2}= \frac{1}{2s}\int_{-s}^s \exp{\big (\imath s'
\omega_1\omega_2\big )}ds'
\end{equation}
and using $(\ref{psi0})$ as zero iteration allow to make integration over time variables $t_1, \ t_2$ in
$(\ref{psiint1b})$ explicitly and get expressions for first $A^{(1)}$, second $A^{(2)},$ etc. iterations:
\begin{eqnarray} \label{iteration1}
A^{(1)}=p^{(1)}\int^{\tilde s}_{-\tilde s}\frac{\exp{\big (
-\frac{\beta}{2}\frac{3\imath+s'}{\imath+3s'}t^2\big )}}{\sqrt{1-2\imath s'+3s'^2}}ds',
  \nonumber \\ A^{(2)}=p^{(2)}\int\int\int\int^{\tilde s}_{-\tilde s}\frac{\exp(-qt^2)}
          {\sqrt{f_1f_2f_3g}}ds'ds_1ds_2ds_3,\\
         \nonumber \\
          A^{(3)}=\ldots \qquad  \qquad \qquad \qquad \qquad\nonumber
\end{eqnarray}
where   $\tilde s=\beta s,$  integration limits for all variables are $[-\tilde s,\tilde s]$, $f_j=1-2\imath s_j+3s_j^2$, $b_j=\frac{\imath+3s_j}{2\beta(3\imath+s_j)},$
$\quad j=1,2,3,$ $g=4(4b_1b_2b_3-4\imath b_1b_2s'+(b_1+b_2+b_3)s'^2),$
$q=(4(b_1b_2+b_1b_3+b_2b_3)+4\imath s'b_3+s'^2)/g,$    and $p^{(n)}$ is determined by the condition
$A^{(n)}|_{t=0}=p$ for all $n.$ Note that instead of keeping $A^{(n)}\big|_{t=0}$  constant for all
iterations it is also possible to  fix the integral $P_0\equiv \int |A^{(n)}|^2 dt$ if one aims to get soliton
solution with some definite value of $P_0$. In principle any values of $\lambda, \beta$ can be chosen for zero iteration
but the most useful is to choose them in such a way to get $A^{(1)}$ as close as
possible to $A^{(0)}$ in order to get faster convergence of iterations to true solution of  eq. $(\ref{psiint1c}).$
It is convenient to make series expansion $A^{(1)}$ in powers of $t^2$ which allows to integrate all terms of this expansion explicitly.
In particular
\begin{eqnarray} \label{beta2}
   p^{(1)}=\sqrt{3}p/\big(arcsinh\frac{3\tilde s-\imath}{2}+c.c.\big),
        \nonumber \\
  A_{ tt}^{(1)}\big|_{t=0}=\\-p^{(1)}\frac{2\beta}{3}\Big(2\sqrt{\frac{\tilde s+\imath}{3\tilde
  s-\imath}}+\frac{\sqrt{3}}{6} arcsinh\frac{3\tilde s-\imath}{2}+c.c. \Big),\nonumber
\end{eqnarray}
where c.c. means complex conjugation. Equating terms proportional $t^2$ in series expansion of
$A^{(1)}$ and $(\ref{psi0})$ one gets from $(\ref{beta2})$ transcendental equation for $\tilde s$ which
gives
\begin{equation} \label{cbeta}
  \tilde s=\beta s=2.393\ldots
\end{equation}
and using $(\ref{psiint1c})$ with the same accuracy
\begin{equation} \label{lamb}
  \lambda =p^2 \cdot 0.482\ldots
\end{equation}
 Fig.1 shows the results of first four iterations obtained
numerically   for $p=1.7,\ d_1=500,\ L=0.01,\ s=2.5,\ \langle d \rangle=0,$  $\tilde t=t/\sqrt{s}$ and $\beta, \ \lambda$ are given by $(\ref{cbeta}),(\ref{lamb}).$
It is enough to consider only positive values of $t$ because $A(t)$ is an even function of $t.$
Note that in case $\langle d \rangle=0$ eq. $(\ref{psiint1c})$ is invariant under two scaling transforms: $t\to \delta_1 t, \
s\to \delta_1^2 s, \ A(\tau)\to A(\tau/\delta_1)$ and $\lambda\to  \delta^2_2 \lambda, \ A\to \delta_2 A$, which allows to extend results of Fig.1 to all values of system parameters $p,\, s$ (of course it has a sense to consider only such values of $p, \, s$ for which condition $Z_{nl}\gg L$ holds).

It is seen in Fig. 1 that the convergence of iteration procedure to soliton solution is very fast for small $t$ and becomes slower for larger
$t$. For $|\tilde t|\stackrel{<}{\sim}1.3$ the renormalization of $A^{(1)}$ by next iteration is very small. In
particular for all $n=2,3,\ldots$ $\ (A_{tt}^{(n)}/A^{(n})\big|_{t=0}$ differs from $\beta$ determined by $(\ref{cbeta})$ for less
that $2\%$. For $|\tilde t|\stackrel{<}{\sim}2.3$ the good approximation is given by second iteration, for  $| \tilde t|\stackrel{<}{\sim}3.2$ by
third etc. The sequence of iteration can be interpreted as evolution of $A$ along some artificial
coordinate $\tilde z$ and it can indicate that the convergence of the solution of $(\ref{psiint1b})$ during evolution along coordinate $z$ to soliton solution  $(\ref{psiint1c})$ should also be slower for larger $t$ what is really observed in
numerical experiments if $\psi|_{z=0}$ is chosen in Gaussian form \cite{mez1}.

Thus, provided the form of soliton is close to Gaussian for small and moderate $t$, the value of $ \beta s$ is universal and is given by $(\ref{cbeta})$.  Zakharov and Manakov \cite{zakhmanak1} proposed the other theory of DM soliton in strong DM limit based on the assumption that typical width of $A$ distribution is much less than $\sqrt{s}.$ In present notation it means $\beta s\gg 1$. Apriori one can not exclude that  soliton solution with $\beta s \gg 1$ is also possible in addition to obtained here. But such solution can not be Gaussian-like and, according to best of my knowledge, in all numerical experiment so far the value of $\beta s$ was of the order of unity (see e.g. \cite{smithknox1,ablowitz1,turmez1}).
Note that choice of  $\beta s \gg 1$ for zero iteration results in convergence of iteration sequence $(\ref{iteration1})$ to Gaussian-like solution with $\beta$ is given by $(\ref{cbeta})$.
It can indicate that solution of $(\ref{psiint1b})$ for $\beta s \gg 1$ also converges  for large $z$ to that Gaussian-like  solution.
One can not also exclude the existence of soliton with  $\beta s \gg 1$ for  $\langle d \rangle\neq 0$ which is outside the scope of the present paper.

Let us try to explore in more details in what sense soliton solution $A$ is close to Gaussian.
From $(\ref{iteration1}),(\ref{beta2})$ and $(\ref{cbeta})$ one can get series expansion (the approximate numerical value of $\tilde s$ is used here to avoid writing a cumbersome expression for every term of expansion):
\begin{eqnarray} \label{A1exp}
 A^{(1)}\exp{\big (\frac{\beta t^2}{2}\big )}/p\simeq1-8.211 \cdot 10^{-2}\ \tilde t^4-2.654 \cdot 10^{-2}\ \tilde t^6\nonumber \\-4.383\cdot 10^{-3}\tilde t^8-
3.946 \cdot 10^{-4}\ \tilde t^{10}+ \ldots
\end{eqnarray}
One can conclude from this expansion that $A$ can be written as
multiplication of Gaussian exponent $(\ref{psi0})$ on a slow function which changes significantly
on scales $\tilde t^2\sim 3.5$. For this and larger scales there is no similarity between soliton and
$(\ref{psi0}).$ Thus DM soliton is close to Gaussian exponent only with accuracy up to numerically
"small parameter" $\sim 1/4$  (because scale of Gaussian exponent is $\beta t^2/2=\tilde s \tilde t^2/2\sim 1$) and there is no really small parameter describing this closeness.
Nevertheless appearance of numerical "small parameter" $1/4$ explains success of expansion of DM
soliton solution in  Hermite polynomials \cite{turmez1,kaup1} which is basically equivalent to expansion
$(\ref{A1exp})$ in Taylor series.

For larger scales $|\tilde t|\stackrel{>}{\sim}1.3$ the correct presentation of DM soliton requires a lot of terms of series
expansion. Instead one can get asymptotical behavior from $(\ref{iteration1})$. E.g. the asymptotic of the first
iteration is given by:
\begin{eqnarray} \label{asymp1}
A|_{t\to \infty}=\frac{p^{(1)}}{16 \beta^2 t^2}\exp{\big (
-\frac{\beta}{2}\frac{3\imath+\tilde s}{\imath+3\tilde s}t^2\big )}\big(\frac{3\tilde s+\imath}{\tilde s-\imath}\big)^{\frac{3}{2}} \times   \nonumber \\
\Big [-4\imath(\tilde s-\imath)+\frac{(3\tilde s-5\imath)(3\tilde s+\imath)}{t^2}\Big]+c.c.+O(\frac{1}{t^6})
\end{eqnarray}
Fig.2 shows that eq. $(\ref{asymp1})$ (curve 2) is a relatively good approximation of $A^{(1)}$ (curve 1) for $|\tilde t|\stackrel{>}{\sim}1.6$.
But of course to improve accuracy of  soliton solution approximation it is necessary to get asymptotic of next iterations. Another way
is to use the  series expansion $(\ref{A1exp})$  as zero iteration instead of
$(\ref{psi0})$. E.g. keeping first four terms of this expansion for zero iteration one can get parametric integral
similar to $(\ref{iteration1})$ which gives good approximation of $A^{(2)}$ in $(\ref{iteration1})$ for $|\tilde t|\stackrel{<}{\sim}3.5.$
Detail consideration of this approximation is outside the scope of the present paper.

The author thanks I.R. Gabitov and E.A. Kuznetsov for helpful discussions.

The support was provided by  the Department
    of Energy, under contract W-7405-ENG-36, RFBR and the program of government support for
leading scientific schools.

E-mail address:  lushnikov@cnls.lanl.gov



\section*{Figure captions:}
~

\noindent Fig.1. Time-dependence of first, second, third and fourth iteration (curves 1,2,3,4 respectively).

\noindent Fig.2. First iteration $A^{(1)}(t)$ (curve 1) versus eq. $(\ref{asymp1})$ (curve 2).
\end{multicols}

\newpage

\begin{figure*}
\centerline{
\epsfxsize=12.2cm
\epsffile{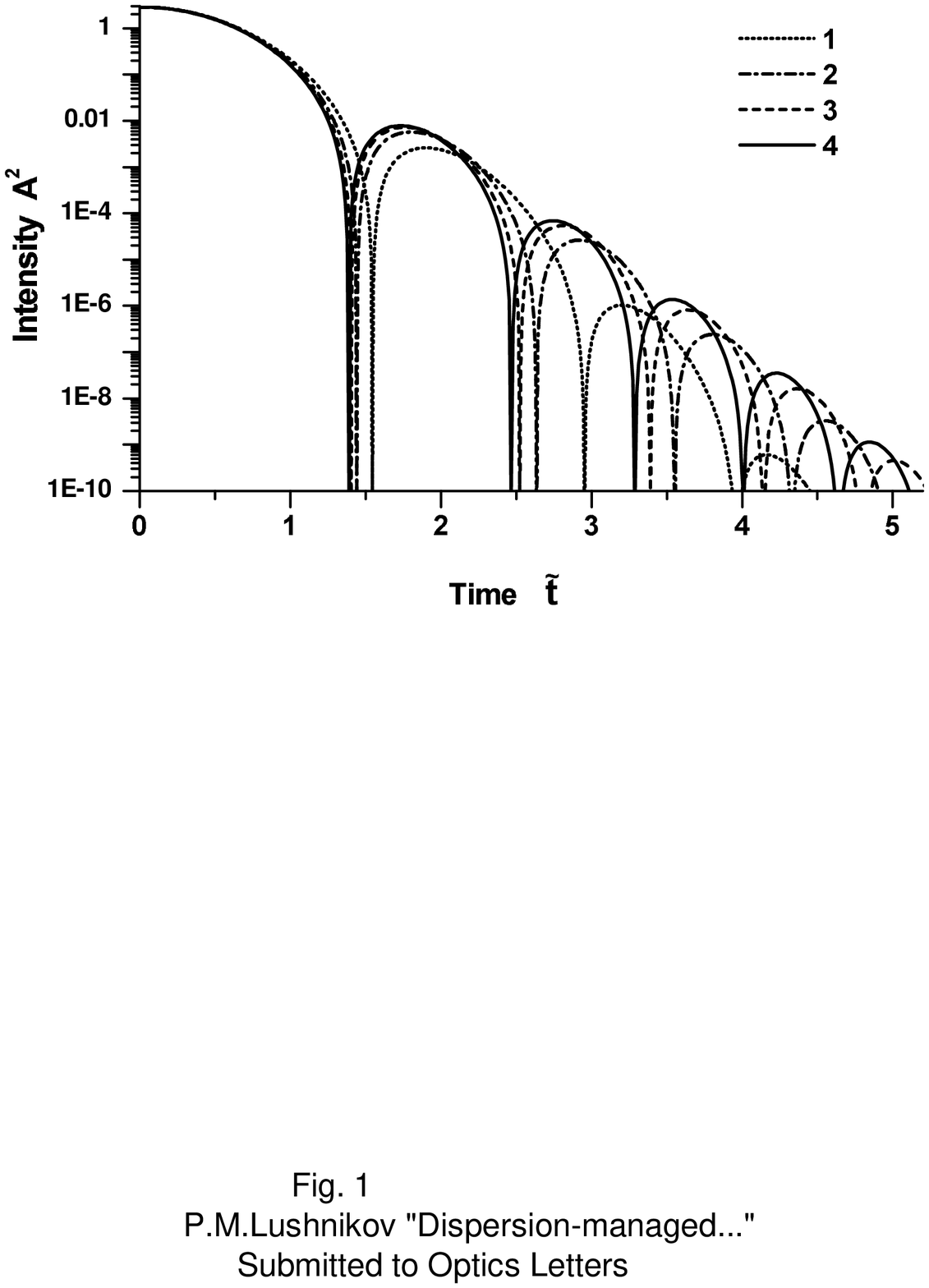}
}
\end{figure*}

\newpage

\begin{figure*}
\centerline{
\epsfxsize=12.2cm
\epsffile{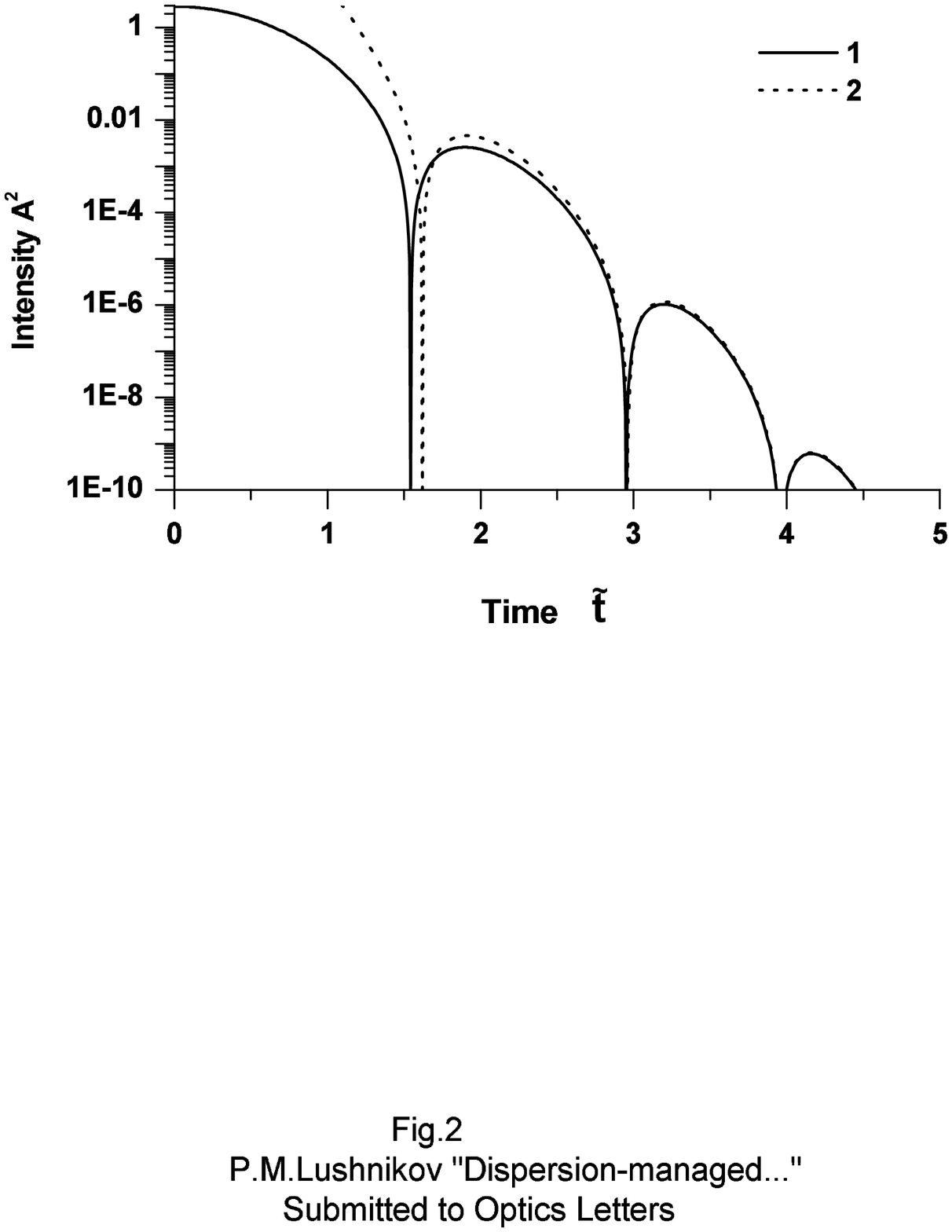}
}
\end{figure*}


\end{document}